# Dynamic detection of current-induced spin-orbit magnetic fields: a phase independent approach

L. Chen[1,2], R. Islinger[2], J. Stigloher[2], M. M. Decker[2], M. Kronseder[2], D. Schuh[2], D. Bougeard[2], D. Weiss[2] and C. H. Back[1]

[1]Department of Physics, Technical University of Munich, Garching b. Munich, Germany.

[2]Institute of Experimental and Applied Physics, University of Regensburg, Regensburg, Germany.

Correspondence and requests for materials should be addressed to L.C and C.H.B. (e-mail: lin0.chen@tum.de, christian.back@tum.de).

**Current induced spin-orbit torques (SOTs) in ferromagnet/non-magnetic metal heterostructures open vast possibilities to design spintronic devices to store, process and transmit information in a simple architecture. It is a central task to search for efficient SOT-devices, and to quantify the magnitude as well as the symmetry of current-induced spin-orbit magnetic fields (SOFs). Here, we report a novel approach to determine the SOFs based on magnetization dynamics by means of time-resolved magneto-optic Kerr microscopy. A microwave current in a narrow Fe/GaAs (001) stripe generates an Oersted field as well as SOFs due to the**

**reduced symmetry at the Fe/GaAs interface, and excites standing spin wave (SSW) modes because of the lateral confinement. Due to their different symmetries, the SOFs and the Oersted field generate distinctly different mode patterns. Thus it is possible to determine the magnitude of the SOFs from an analysis of the shape of the SSW patterns. Specifically, this method, which is conceptually different from previous approaches based on lineshape analysis, is phase independent and self-calibrated. It can be used to measure the current induced SOFs in other material systems, e.g., ferromagnetic metal/non-magnetic metal heterostructures.**

The investigation of the mutual conversion between charge and spin currents has witnessed significant attention in recent years due to its possible technological impact for spintronic devices[1,2]. In ferromagnet (FM)/non-magnetic metal (NM) heterostructures, a charge current flowing in the NM along the *x*-axis will generate a transverse spin accumulation $\sigma$ along the *y*-direction at the interface via the spin Hall effect and/or the inverse spin galvanic effect[1]. The resulting spin accumulation acts on the ferromagnetic layer via field-like ($\tau_{FL}$) and damping-like ($\tau_{DL}$) spin-orbit torques (SOTs), which can be written as $\tau_{FL} = -\gamma\mu_0 h_{FL} \mathbf{m} \times \mathbf{y}$ and $\tau_{DL} = -\gamma\mu_0 h_{DL} \mathbf{m} \times \mathbf{m} \times \mathbf{y}$, where $\gamma$ is the gyromagnetic ratio, $\mu_0$ the magnetic constant, $\mathbf{m}$ the magnetization unit vector, and $h_{FL}$ ($h_{DL}$) the corresponding effective field-like (damping-like) spin-orbit magnetic field $\mathbf{h}_{SOF}$. These torques modify the magnetization's equation of motion, i.e., the Landau-Lifshitz-Gilbert (LLG) equation and are responsible for a number of spin-orbit related functionalities including magnetization switching[3,4], domain wall motion[5-

[7], and auto-oscillations of the magnetization[8,9].

## Historic development of the spin-transfer torque ferromagnetic resonance method

To optimize material parameters leading to efficient SOTs, the magnitude of the SOFs must be determined accurately. One frequently used approach is the spin-transfer-torque ferromagnetic resonance (STT-FMR) method[10], which is based on a lineshape analysis of the rectified d.c. voltage induced by FMR. It is generally assumed, that the symmetric component of the d.c. voltage, $V_{sym}$, corresponds to the out-of-plane $h_{DL}$ while the anti-symmetric component, $V_{a-sym}$, to the in-plane Oersted field generated by the current flowing in NM (see Supplementary Note 1 for details). This method is so-called self-calibrated since the spin Hall angle in the NM (related to $h_{DL}$) is determined by the ratio of $V_{sym}/V_{a-sym}$. Initially, the importance of $h_{FL}$, which also generates an anti-symmetric voltage signal $V_{a-sym}$, has not been properly taken into consideration. Pai *et al*. further modified this method and extracted $h_{DL}$ and $h_{FL}$ by measuring the dependence of $V_{sym}/V_{a-sym}$ on the FM layer thickness $t_{FM}$, assuming that $h_{FL}$ is independent of $t_{FM}$[11,12]. However, this does not hold since magneto-optical[13] and magneto-transport[14] methods show that $h_{FL}$ strongly depends on $t_{FM}$, which possibly leads to a wrong estimation of $h_{FL}$ and $h_{DL}$. A second well-established technique based on FMR is the spin-orbit-torque FMR (SOT-FMR) method, which has been utilized to characterize the SOFs in single crystalline ferromagnetic materials with broken inversion symmetry[15]. In contrast to bi-layer systems, there is no in-plane Oersted field since only one layer is involved, and the single crystalline ferromagnet acts both as spin current generator and detector (see

Supplementary Note 1 for the differences between STT-FMR and SOT-FMR). Up to now, STT-FMR and SOT-FMR have been used to study spin-orbit related phenomena in a large variety of materials (see the large number of references which cite Refs. 10 and 15), including non-magnetic metals[4,16], topological materials[17-20], magnetic semiconductors[21], antiferromagnets[22-24] and transition-metal dichalcogenides[25-28]. It should be noted that for both, STT-FMR and SOT-FMR, the out-of-plane Oersted field $h_{\text{rf}}^{\text{FM},z}$ generated by the current flowing in the ferromagnetic material itself contributes no net effect to the measured d.c. voltage since it is anti-symmetrically distributed (see Supplementary Note 1). This, however, becomes the key ingredient for the present study.

For electric-current driven FMR, it is generally believed that the magnetization dynamics **m**($t$) and the microwave driving current **j**$_{\text{rf}}$ are coherently coupled and there is no phase difference $\varphi_{j\text{-}m}$ between these dynamic quantities, i.e., $\varphi_{j\text{-}m} = 0$ always holds. Only recently, it has been noticed by spatially resolved ferromagnetic resonance phase imaging that a possible phase difference transverse to a CoFeB/Pt stripe exists[29], and affects the determination the magnitude of $h_{\text{DL}}$. Note that for the sample studied in Ref. 29, a part of $j_{\text{rf}}$ flows also in the CoFeB layer, and the generated $h_{\text{rf}}^{\text{FM},z}$ can influence the lineshape and subsequently the phase. Therefore, the open questions, which have not been properly addressed so far, are: does $\varphi_{j\text{-}m} = 0$ always hold for any spin-current-source material, irrespective of the details of the material/device? If not, is it still possible to quantitatively determine SOFs by magnetization dynamics?

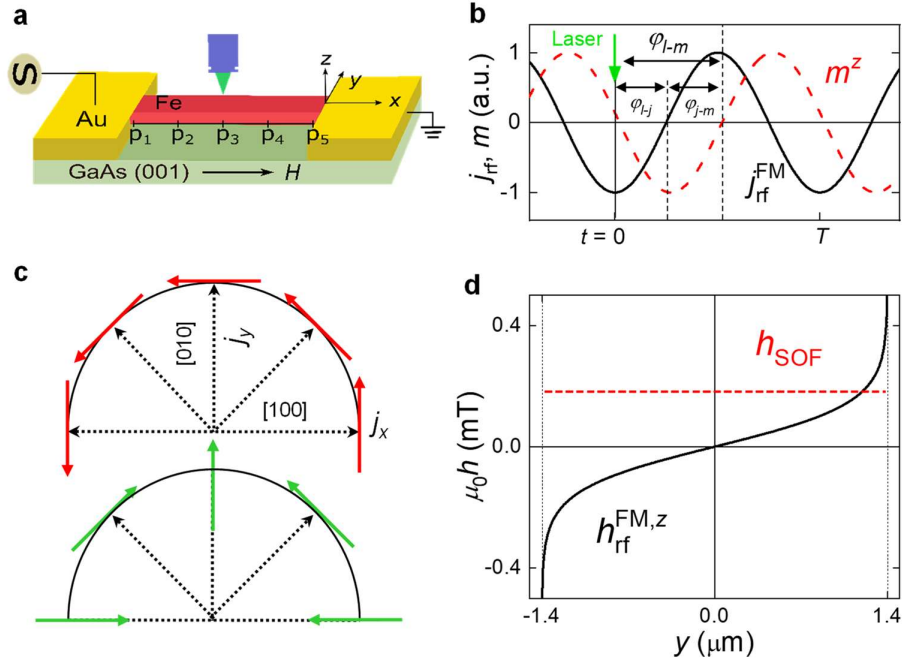

**Figure 1 | Schematic of device and driving fields. a**, Schematic of the device used for the detection of magnetization dynamics driven by electric-current. The out-of-plane component of the dynamic magnetization $m^z(t)$ is detected by time-resolved magneto-optical Kerr (TRMOKE) microscopy. A microwave current $j_{rf}^{FM}$ with a frequency $f$ of 12 GHz is fed into the Fe stripe deposited on a semi-insulating GaAs(001) substrate, and excites $\mathbf{m}(t)$ by the combination of spin-orbit field $h_{SOF}$ and Oersted field $h_{rf}^{FM,z}$. The external magnetic field $H$ is applied parallel to $j_{rf}^{FM}$. **b**, Phase relations in the TRMOKE set up. $\varphi_{l-j}$ is the phase difference between the laser pulse (green arrow) and $j_{rf}^{FM}$ (black line), which is a controlled parameter. $\varphi_{j-m}$ is the assumed phase difference between $j_{rf}^{FM}$ and $m^z(t)$ (red dashed line). The phase difference between the laser pulse and $m^z(t)$, $\varphi_{l-m}$, is thus the sum of $\varphi_{l-j}$ and $\varphi_{j-m}$, i.e., $\varphi_{l-m} = \varphi_{l-j} + \varphi_{j-m}$. **c**, Current-orientation dependence of $h_{SOF}$ induced by Bychkov-Rashba-like (red) and Dresselhaus-like (green) spin-orbit

interaction. Since H is parallel to $j_{rf}^{FM}$, only the transverse components of $h_{SOF}$ excite magnetization dynamics. **d**, Lateral distributions of $h_{SOF}$ (red dashed line) and $h_{rf}^{FM,z}$ (black line). $h_{SOF}$ is symmetric across y, while $h_{rf}^{FM,z}$ is anti-symmetric. The different symmetries of the excitations lead to distinctive standing spin wave patterns, i.e., the symmetric $h_{SOF}$ excites even spin wave modes (n = 1, 3…), while the anti-symmetric $h_{rf}^{FM,z}$ excites odd modes (n = 2, 4 …).

## Results
### Evidence of phase shift for electric-current driven FMR

Here, we use Fe/GaAs (001) bi-layers as a model material system to investigate a possible variation of $\varphi_{j-m}$, see Fig. 1a. The advantages of the single crystalline system Fe/GaAs are: i) presence of sizeable interfacial SOFs having the same symmetry as FM/NM bi-layers[30]; ii) low Gilbert damping constant, iii) the electrical current $j_{rf}^{FM}$ flows solely in Fe and thus a complex analysis can be avoided (see Supplementary Note 1); iv) tuneable resistivity of Fe simply by changing the Fe layer thickness $t_{Fe}$. The measurements are carried out by phase-sensitive time-resolved magneto-optical Kerr effect (TR-MOKE) microscopy[31] (see Methods). As shown in Fig. 1b, the phase difference between the pulse laser and the dynamic magnetization **m**(t), $\varphi_{l-m}$, can be written as,

$$\varphi_{l-m} = \varphi_{l-j} + \varphi_{j-m}, \quad (1)$$

where $\varphi_{l-j}$ is the controlled phase between pulse laser and $j_{rf}^{FM}$, and $\varphi_{j-m}$ the assumed phase difference between $j_{rf}^{FM}$ and **m**(t). The polar Kerr signal at a certain $\varphi_{l-j}$, $V_{Kerr}(\varphi_{l-j})$, is proportional to the real part of the out-of-plane component of the dynamic magnetization $m^z$, which can be obtained from the complex dynamic susceptibility[30]

$$V_{\text{Kerr}}(\varphi_{l\text{-}j}) \sim [\text{Re}(\chi^o)h^o - \text{Im}(\chi_a^i)h^i]\cos\varphi_{l\text{-}m} - [\text{Im}(\chi^o)h^o + \text{Re}(\chi_a^i)h^i]\sin\varphi_{l\text{-}m}. \qquad (2)$$

Here $\text{Re}(\chi^o)$ [$\text{Im}(\chi^o)$] is the real (imaginary) part of the diagonal dynamic susceptibility due to out-of-plane excitation $h^o$, and $\text{Re}(\chi_a^i)$ [$\text{Im}(\chi_a^i)$] the real (imaginary) part of the off-diagonal dynamic susceptibility due to in-plane excitation $h^i$. As shown in Fig. 1d, for Fe/GaAs studied here, $h^i$ contains only the position independent SOF $h_{\text{SOF}}{}^y$, i.e., $h^i = h_{\text{SOF}}{}^y$. In contrast $h^o$ contains both the position dependent Oersted field $h_{\text{rf}}^{\text{FM},z}$ and a position independent SOF $h_{\text{SOF}}{}^z$, i.e., $h^o = h_{\text{rf}}^{\text{FM},z}(y) + h_{\text{SOF}}{}^z$. It is worth mentioning that $\text{Im}(\chi_a^i)$ and $\text{Im}(\chi^o)$ [$\text{Re}(\chi_a^i)$ and $\text{Re}(\chi^o)$] show symmetric (anti-symmetric) lineshape with respect to the external magnetic field $H$, and their magnitudes can be calculated by solving the LLG equation[30].

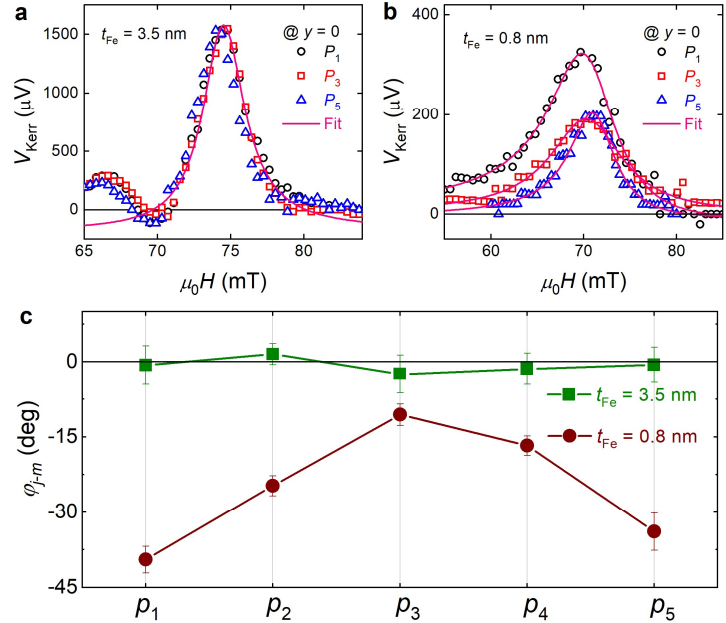

**Figure 2 | Determination of $\varphi_{j\text{-}m}$ for electric-current driven magnetization dynamics.** Position dependent Kerr voltage $V_{\text{Kerr}}$ measured at the center of the stripe ($y$

= 0, where $h_{\text{rf}}^{\text{FM},z}=0$) for **a** Fe thickness $t_{\text{Fe}}=3.5$ nm and **b** $t_{\text{Fe}}=0.8$ nm. For both devices, $\varphi_{l\text{-}j}$ is set to 50° and the device dimension is 6.4 μm × 100 μm. One can see that both the magnitude and the lineshape for $t_{\text{Fe}}=3.5$ nm remain unchanged but change dramatically for $t_{\text{Fe}}=0.8$ nm. The solid lines in **a** and **b** are fits to equation (3), which give the magnitude of $\varphi$. The bump at about 66 mT for $t_{\text{Fe}}=3.5$ nm is due to the formation of a standing spin wave, see discussion in below. **c,** Position dependence of $\varphi_{j\text{-}m}$ obtained from equation (4), displaying a clear variation of $\varphi_{j\text{-}m}$ for $t_{\text{Fe}}=0.8$ nm.

Figs. 2a and b show the position dependence of the Kerr voltage $V_{\text{Kerr}}$ measured at the centre of the stripe ($y=0$ and $h_{\text{rf}}^{\text{FM},z}=0$) under $\varphi_{l\text{-}j}=50°$ for $t_{\text{Fe}}=3.5$ nm and 0.8 nm. Both devices have the same dimension of 6.4 μm × 100 μm but show an opposite temperature coefficient in the temperature dependence of the resistivity (see Supplementary Note 2). For $t_{\text{Fe}}=3.5$ nm, the lineshape as well as the magnitude of $V_{\text{Kerr}}$ remains the same along the stripe from $p_1$ to $p_5$, while they change dramatically for $t_{\text{Fe}}=0.8$ nm. To extract $\varphi_{j\text{-}m}$, the characteristic $V_{\text{Kerr}}$ spectra can be fitted by

$$V_{\text{Kerr}}=A\frac{\cos\varphi\Delta H^2+\sin\varphi\Delta H(H-H_R)}{4(H-H_R)^2+\Delta H^2}. \tag{3}$$

Here $A$ is an apparatus dependent coefficient, $H_R$ the magnetic field at FMR, $\Delta H$ the full width at half maximum, and $\varphi$ is the phase factor which determines the lineshape of $V_{\text{Kerr}}(H)$. From equations (1) to (3), the magnitude of $\varphi_{j\text{-}m}$ can be derived as

$$\varphi_{j\text{-}m}=\tan^{-1}\frac{\text{Re}(\chi^o)h^o\cos\varphi+\text{Im}(\chi_a^i)h^i\sin\varphi}{\text{Re}(\chi_a^i)h^i\cos\varphi-\text{Im}(\chi^o)h^o\sin\varphi}-\varphi_{l\text{-}j}, \tag{4}$$

which provides a measure of $\varphi_{j\text{-}m}$ via TRMOKE spectra. Using the corresponding dynamic susceptibilities as well as $\varphi$ values for both devices, and considering that $h^o \sim h^i$, the magnitude of $\varphi_{j\text{-}m}$ can be obtained from equation (4). Fig. 2c summarizes $\varphi_{j\text{-}m}$ as a function of position for both devices. One can see that $\varphi_{j\text{-}m}$ remains constant within experimental error for $t_{Fe}$ = 3.5 nm. However, a sizeable variation of $\varphi_{j\text{-}m}$ is observed for $t_{Fe}$ = 0.8 nm. The variation of $\varphi_{j\text{-}m}$ could be due to the fact that the rf characteristics of Fe changes from a good conductor to a dielectric upon decreasing $t_{Fe}$ (see Supplementary Note 2). Since $\varphi_{j\text{-}m}$ influences the lineshape as well as the amplitude of the dynamic magnetization, the variation of $\varphi_{j\text{-}m}$ casts doubt on the legitimacy of using electrical measurements for the quantitative determination of SOFs in more resistive materials (see Supplementary Note 3).

To overcome this problem, it is of vital importance to establish a phase-independent technique to reliably determine the SOFs based on magnetization dynamics. Here, we report a self-calibrated and phase independent approach to measure current induced SOFs by analyzing the shape of the standing spin wave (SSW) mode patterns, i.e., a method which is distinctly different from previous electrical methods based on lineshape analysis.

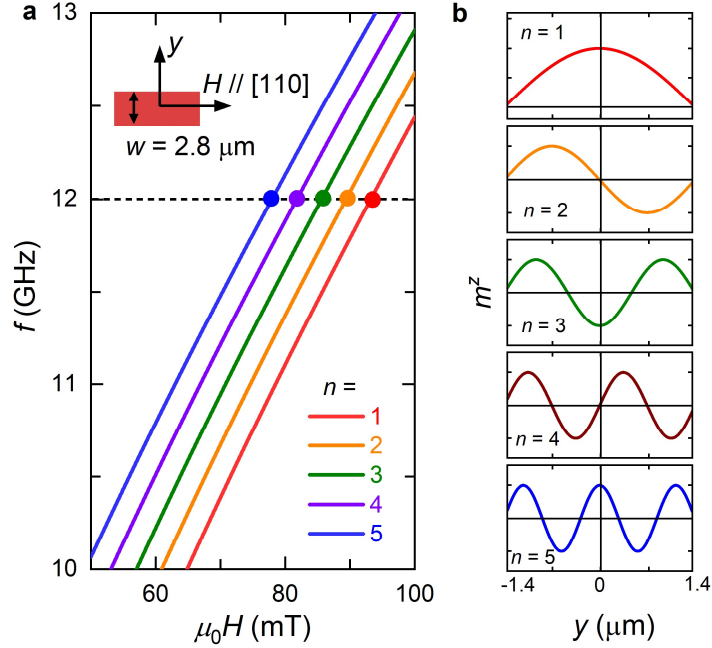

**Figure 3 | Eigenmodes and distribution of confined SSWs. a**, Calculated eigenmodes for a laterally confined Fe/GaAs stripe with $w = 2.8$ μm. The external magnetic field $H$ is applied along the [110]-direction of GaAs, and the intersection defines the required $H_R$ for each standing spin wave (SSW) mode at $f = 12$ GHz. **b**, The lateral distribution of SSW modes for $n = 1$ - 5. The symmetric modes ($n = 1, 3$ and $5$) can be excited by symmetric excitations; anti-symmetric modes ($n = 2$ and $4$) can be excited by anti-symmetric excitations.

### Formation of standing spin waves in a laterally confined Fe/GaAs stripe

Formation of SSW is a prerequisite for this work. Fig. 3a shows the calculated SSW eigenmodes for a 2.8 μm wide, 3.5 nm thick Fe stripe with $H$ applied along the [110]-direction of the GaAs substrate, which corresponds to the Damon-Eshbach geometry[32,33]. In the calculation, the following parameters determined by separate magnetization and FMR measurements are used: saturation magnetization $\mu_0 M_S = 2.1$

T, effective demagnetization field $\mu_0 H_K$ = 1.75 T, and Landé g-factor $g$ = 2.12. The intersection at a frequency $f$ of 12 GHz specifies $H_R$ of each mode, which is expected to be observed in the experiment. The lateral confinement leads to a mode separation of 4 mT (i.e. 8 mT between odd modes), which is comparable to the magnitude of $\Delta H$. The normalized profiles of $m^z$ for the first five modes ($n$ = 1 - 5), i.e., $m^z$ as a function of space coordinate $y$, are displayed in Fig. 3b. One can see that the odd (even) modes are symmetrically (anti-symmetrically) distributed with respect to the center of the stripe. Consequently, the odd (even) modes can be excited by symmetrically (anti-symmetrically) distributed driving fields due to symmetry reasons[32,33].

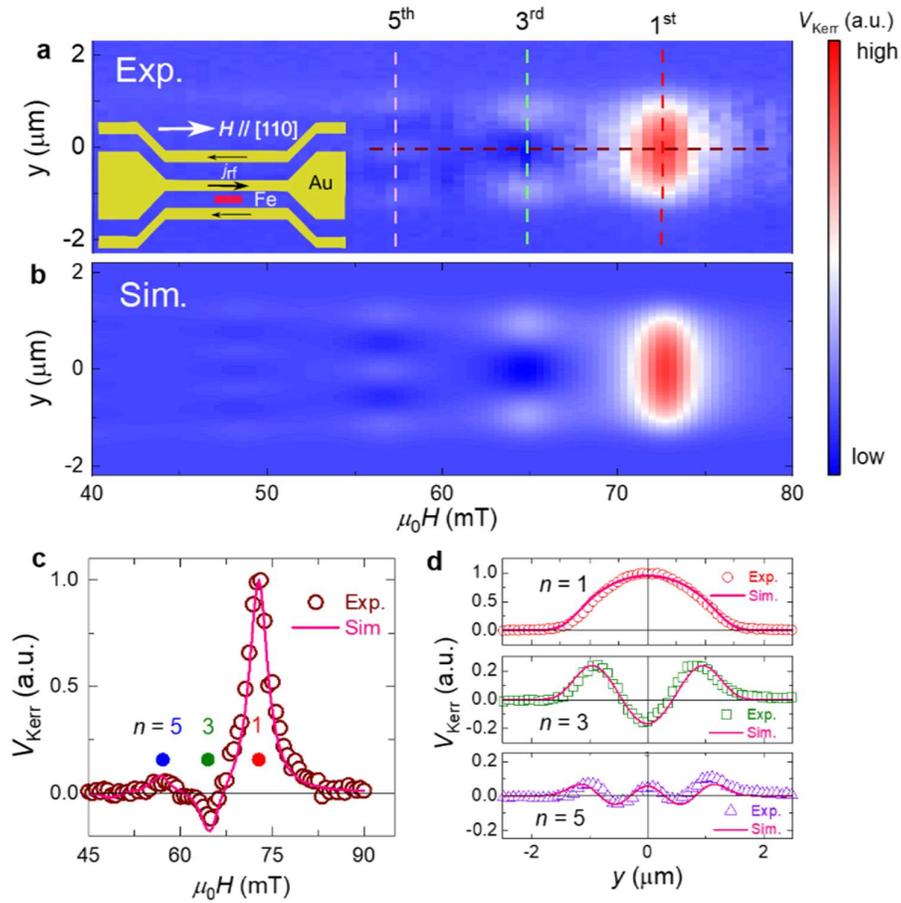

**Figure 4 | SSWs driven by a symmetric excitation. a** SSWs detected in a Fe/GaAs

stripe by TR-MOKE microscopy, where the magnetization dynamics is excited by a homogeneous (symmetric) out-of-plane Oersted field through a coplanar waveguide (CPW). Only symmetric odd modes ($n$ = 1, 3 and 5) can be observed. The inset shows the schematic of the CPW device, where the Fe stripe is integrated into the gap of the CPW (for details see Methods), and $H$ is applied along the long axis of the stripe, i.e. along the [110]-direction of GaAs. **b**, Micromagnetic simulation of the SSW modes using MuMax3, which reproduces the experimentally observed modes well. In the simulation, we use the same material parameters as for the calculation of the dispersion relation and we convolve the simulation with a Gaussian beam profile (see Supplementary Note 3). **c**, Horizontal line cut of the Kerr signal at the center of the stripe ($y$ = 0). The three peaks can be fitted by symmetric Lorentzians, and the positions of the 1$^{st}$, 3$^{rd}$ and 5$^{th}$ mode are indicated by red, green and blue circles, respectively. **d**, Vertical cut of modes for $n$ = 1, 3 and 5. All the modes show symmetric profiles and can be well fitted by MuMax simulations.

We first analyse the eigenmodes of the Fe/GaAs stripe under homogeneous (symmetric) excitation. The stripe, which is 2.8 μm in width and 20 μm in length with the long side along the [110]-direction of the GaAs substrate, is integrated in the gap of a coplanar waveguide (CPW) by using electron-beam lithography, as shown in the inset of Fig. 4a (see Methods). Here, the Fe stripe is exposed to homogeneous excitation by an out-of-plane Oersted field $h_{Oe}^z$, which is generated by $j_{rf}$ flowing in the signal and ground line of the CPW. According to equation (2), the detected Kerr signal can be simplified as, $V_{Kerr}(\varphi_{l-j}) \sim \text{Re}(\chi^o)h^z_{Oe}\cos(\varphi_{l-j}+\varphi_{j-m}) - \text{Im}(\chi^o)h^z_{Oe}h^o\sin(\varphi_{l-j}+\varphi_{j-m})$. Fig. 4a shows the normalized $V_{Kerr}(H, y)$ image measured at $\varphi_{l-j} = 90°$. As expected, only the odd modes with $n = 1, 3$ and $5$ appear. Fig. 4b presents the micromagnetic simulation of the SSW modes using the same parameters as those used in Fig. 3, which reproduces the experimental results well (see Supplementary Note 3 for MuMax3 simulations). To have a closer look at the obtained modes, we perform a horizontal scan for the data in Fig. 4a, i.e., by placing the laser at the center of the stripe and sweeping $H$. As shown in Fig. 4c, the cut shows only symmetric lineshapes, which can be fitted using the corresponding cut of the simulation data in Fig. 4b. The locations of the 1st, 3rd and the 5th mode are marked by solid points, and the mode spacing coincides well with the eigenmode calculation shown in Fig. 3a. Note that the mode position differs between Figs. 3(a) and 4(c); this is because the in-plane biaxial and uniaxial magnetic anisotropies are not included in the mode calculation. Since only purely symmetric lineshapes are observed, one can infer that $\varphi_{j-m} = 0°$. Otherwise an anti-symmetric component in the $V_{Kerr}$ trace originating from $\text{Re}(\chi^o)$ is expected. This is not surprising since $j_{rf}$ and $h_{Oe}^z$ are intrinsically in phase due to the fact that the CPW is impedance matched with the rf network. These results also prove the validity of the proposed phase analysis presented above. Fig. 4d shows the 1st, 3rd and 5th mode as a function of lateral space coordinate $y$. All the modes show symmetric profiles with the peak wave-amplitude ratio of ~ 10:2:1, which can also be well fitted by micromagnetic simulations.

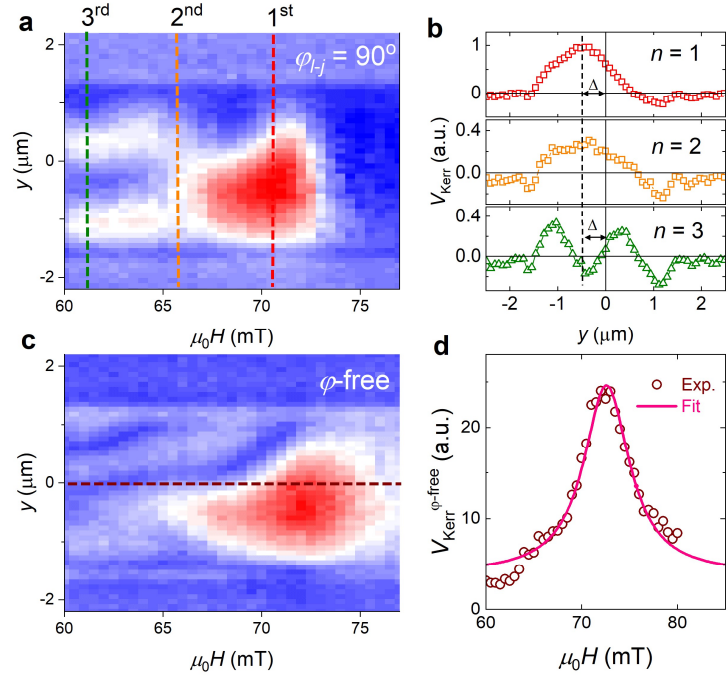

**Figure 5 | SSWs driven by electric-current. a**, Image of the TR-MOKE signal measured at $\varphi_{l\text{-}j} = 90°$ and $\mathbf{j}_{rf}^{FM}$ // **H** // [110]. **b**, Profiles of the first three modes, i.e., vertical cuts along the dashed lines in **a**. The position of the maxima (minima) of $n = 1$ ($n = 3$) shifts away from the center of the stripe by $\Delta \sim 0.4$ μm due to the interference with the 2$^{nd}$ mode, as indicated by the dashed line. **c**, Corresponding image of $V_{Kerr}^{\varphi\text{-free}}$ for a [110]-device obtained by square and root operation of $V_{Kerr}(0°)$ and $V_{Kerr}(90°)$, i.e., $V_{Kerr}^{\varphi\text{-free}} = \sqrt{[V_{Kerr}(0°)]^2 + [V_{Kerr}(90°)]^2}$. **d**, Horizontal cut of $V_{Kerr}^{\varphi\text{-free}}$ at $y = 0$, which can be fitted by a symmetric Lorentzian.

## Determination of SOFs by the shape of the standing spin wave pattern

Next, measurements are performed on a 2.8 μm wide, 100 μm long stripe orientated along the [110]-direction of GaAs using an electric-current excitation as shown in Fig. 1a. A rf-current density $j_{rf}^{FM} = 1.0 \times 10^{11}$ Am$^{-2}$ is applied to the device, and $H$ is set parallel to $j_{\text{rf}}^{\text{FM}}$. The magnitude of $j_{\text{rf}}^{\text{FM}}$ is calibrated by the Joule heating induced resistance increase[15]. As shown in Fig. 1d, the driving fields here contain both symmetric $h_{\text{SOF}}$ and anti-symmetric $h_{\text{rf}}^{\text{FM},z}$ components. In addition to the odd modes driven by the symmetric SOFs, even modes excited by the anti-symmetric $h_{\text{rf}}^{\text{FM},z}$ are expected. Fig. 5a shows the SSW pattern measured at $\varphi_{l-j} = 90°$. In contrast to Fig. 4a where only the symmetric odd modes are observed, for the case of electric-current excitation, both the 1st and 3rd modes are not located at the center of the stripe anymore. This indicates the emergence of the anti-symmetric 2nd mode. Because the mode spacing is of the same magnitude as the FMR linewidth, the nearest modes merge, and the shape of the SSW pattern is dramatically altered and shifted. For example, the 2nd mode increases $V_{\text{Kerr}}$ of the 1st mode on the lower part of the stripe while reducing it on the upper side. A clearer shift of the patterns can be seen from the profile (vertical cut) of each mode. As shown in Fig. 5b, the maximum (minimum) position of the 1st (3rd) mode shifts away from center to the lower part of stripe by an absolute value of $\Delta \sim 0.4$ µm.

If the phase term $\varphi_{j-m}$ is unknown, it is impossible to extract the magnitude of SOF from Fig. 5a. However, it is possible to eliminate $\varphi_{j-m}$ through square and root operations of $V_{\text{Kerr}}(\varphi_{l-j})$ measured at two phases with 90° phase shift. Based on equation (2), the $\varphi_{j-m}$ independent Kerr voltage $V_{\text{Kerr}}^{\varphi\text{-free}}$ can be obtained as

$$V_{\text{Kerr}}^{\varphi\text{-free}} = \sqrt{\left[V_{\text{Kerr}}(\varphi_{l-j})\right]^2 + \left[V_{\text{Kerr}}(\varphi_{l-j}+90°)\right]^2}$$

$$\sim \text{Im}(\chi_a^i) h_{\text{SOF}}^y \sqrt{\left[1 - \frac{\text{Re}(\chi^o)}{\text{Im}(\chi_a^i)} \frac{h^o}{h_{\text{SOF}}^y}\right]^2 + \left[\frac{\text{Re}(\chi_a^i)}{\text{Im}(\chi_a^i)} + \frac{\text{Im}(\chi^o)}{\text{Im}(\chi_a^i)} \frac{h^o}{h_{\text{SOF}}^y}\right]^2}. \quad (5)$$

The corresponding $V_{\text{Kerr}}^{\varphi\text{-free}}$ image for the [110]-oriented device is shown in Fig. 5c. For the present sample with $\mu_0 H_K$ of 1.75 T, the magnitude of the susceptibility under out-of-plane excitation is much smaller than the in-plane one, and the ratios of the dynamic susceptibilities under the square root are determined as[30]: $\text{Re}(\chi^o)/\text{Im}(\chi_a^i) = 0.1$, $\text{Re}(\chi_a^i)/\text{Im}(\chi_a^i) = -0.5$, and $\text{Im}(\chi^o)/\text{Im}(\chi_a^i) = -0.2$. At the center of the stripe ($h_{rf}^z = 0$ and $h^o = h_{SOF}^z$), equation (5) can be further simplified to:

$$V_{\text{Kerr}}^{\varphi\text{-free}} \approx \text{Im}(\chi_a^i) h_{SOF}^y \sqrt{1 + \left[\frac{\text{Re}(\chi^o)}{\text{Im}(\chi_a^i)}\right]^2 + 2\left[\frac{\text{Re}(\chi_a^i)}{\text{Im}(\chi_a^i)} \cdot \frac{\text{Im}(\chi^o)}{\text{Im}(\chi_a^i)} - \frac{\text{Re}(\chi^o)}{\text{Im}(\chi_a^i)}\right] \frac{h_{SOF}^z}{h_{SOF}^y}} = \text{Im}(\chi_a^i) h_{SOF}^y \sqrt{1 + \left[\frac{\text{Re}(\chi^o)}{\text{Im}(\chi_a^i)}\right]^2}.$$

This means only $h_{SOF}^y$ contributes to $V_{\text{Kerr}}^{\varphi\text{-free}}$, and the effect of $h_{SOF}^z$ can be neglected in the analysis due to the large effective demagnetization field. Equation (5) also suggests that, at the centre of the stripe, the lineshape of the $V_{\text{Kerr}}^{\varphi\text{-free}}$ trace is symmetric with respect to $H$, which is confirmed by the horizontal cut shown in Fig. 5d. However, when the laser is moved away from the centre of the stripe, the above assumption becomes invalid, since $h_{rf}^{FM,z} > h_{SOF}^y$ holds. The appearance of even modes excited by $h_{rf}^{FM,z}$ can alter the shape of the odd mode pattern, which, therefore, provides a phase-independent way to determine the magnitude of $h_{SOF}^y$.

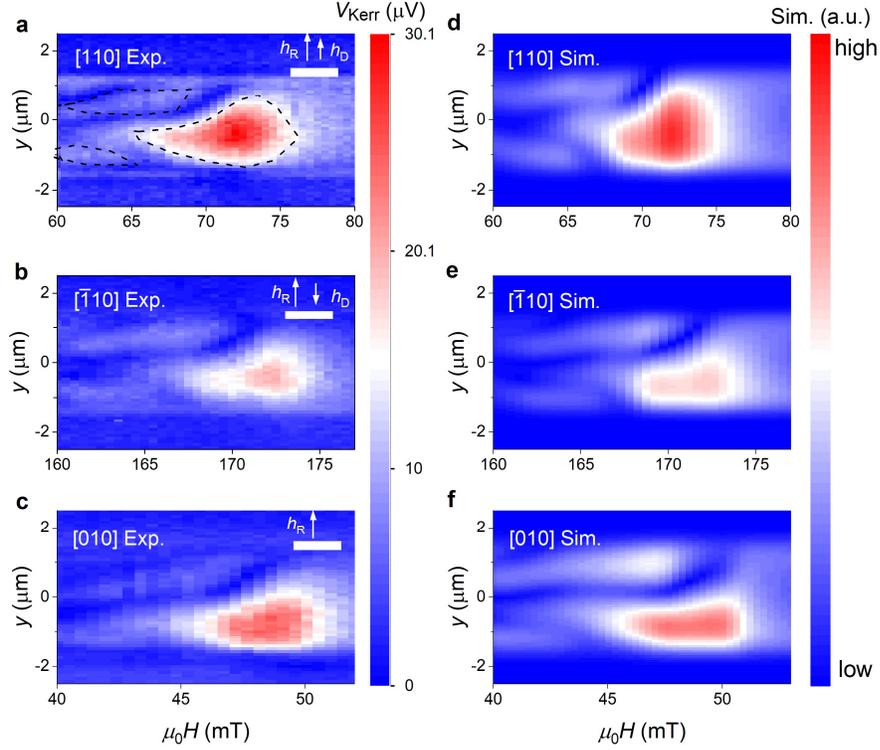

**Figure 6 | Determination of SOF by the shape of SSW pattern.** Image of $V_{\text{Kerr}}^{\varphi\text{-free}}(H,y)$ signal for **a** $j_{\text{rf}}^{\text{FM}}$ // **H** // [110], **b** $j_{\text{rf}}^{\text{FM}}$ // **H** // [$\bar{1}$10] and **c** $j_{\text{rf}}^{\text{FM}}$ // **H** // [010]. In the plots, $j_{\text{rf}}^{\text{FM}}$ has been normalized to $1 \times 10^{11}$ Am$^{-2}$. The configurations of the SOFs induced by Bychkov-Rashba-like $h_R$ and Dresselhaus-like $h_D$ are also presented in the insets. $h_R$ and $h_D$ are constructive for [110]-oriented devices, but destructive for [$\bar{1}$10]-oriented devices. For [010]-orientation, only $h_R$ is detected. Corresponding images obtained by micromagnetic simulations for devices oriented along **d** [110], **e** [$\bar{1}$10] and **f** [010].

Figs. 6a-c present the images of $V_{\text{Kerr}}^{\varphi\text{-free}}(H,y)$ for devices structured along [110]-, [$\bar{1}$10]- and [010]-orientations. To compare the amplitudes of $V_{\text{Kerr}}^{\varphi\text{-free}}$, all images are normalized to the current density $j_{\text{rf}}^{\text{FM}} = 1 \times 10^{11}$ Am$^{-2}$. As shown in the images, the coalescence of the first three mode patterns leads to the formation of three main regions as indicated by the closed dashed lines. The odd and even modes merge and become

indistinguishable after the treatment of square and root operations. All the $V_{\text{Kerr}}^{\varphi\text{-free}}(H, y)$ images show similar patterns indicating similar excitations for each device. However, the maximum intensity of the Kerr signal, $V_{\max}$, differs significantly for different crystallographic directions with $V_{\max}^{[110]} = 1.2 V_{\max}^{[010]} = 1.7 V_{\max}^{[\bar{1}10]}$. This implies a dependence of $h_{\text{SOF}}{}^y$ on the current direction due to interference of Bychkov-Rashba-like and Dresselhaus-like spin-orbit interactions. As sketched in the insets of Figs. 6a to c, constructively aligned Dresselhaus $h_D$ and Bychkov-Rashba $h_R$ SOFs are detected ($h_{\text{SOF}}{}^y = h_R + h_D$) for the [110]-orientation, while for the [$\bar{1}$10]-orientation, $h_D$ and $h_R$ add destructively ($h_{\text{SOF}}{}^y = h_R - h_D$). For the [010]-orientation only $h_R$ can be detected ($h_{\text{SOF}}{}^y = h_R$). To quantify $h_D$ and $h_R$, we repeat the micromagnetic simulations, similar to the case where the magnetization dynamics is only driven by a homogeneous $h_{\text{Oe}}^z$ originating from the CPW, but now including both $y$-independent $h_{\text{SOF}}{}^y$ and $y$-dependent $h_{\text{rf}}^{\text{FM},z}$ calculated from Biot-Savart's law. A least square algorithm is carried out to minimize the difference between images obtained by experiment and simulations (see Supplementary Note 3 for the fitting details). As shown by Figs. 6d to f, the corresponding simulation images reproduce the experiments reasonably well. For the [110]-device, the magnitude of the extracted SOF, $\mu_0 h_R + \mu_0 h_D = 0.28$ mT; and for the [$\bar{1}$10]-device, $\mu_0 h_R - \mu_0 h_D = 0.13$ mT, which gives $\mu_0 h_R = 0.21$ mT and $\mu_0 h_D = 0.07$ mT. The magnitude $\mu_0 h_R$ in turn is consistent with the value determined from a [010]-device with $\mu_0 h_R = 0.22$ mT, indicating the validity of our method. Although the present experiment only determines the magnitude of the field-like torque (corresponding to $h_{\text{SOF}}{}^y$) due to the relatively large $H_K$ value, we propose in the Supplementary Note 4 that it is also possible to determine the magnitude of field-like and damping-like torques in FM/NM bi-layers with a reduced $H_K$, showing the completeness of this method.

**Conclusions**

We have demonstrated by TR-MOKE measurements that a possible phase variation can be detected when using electric-current excitation. Although electrical detection of magnetization dynamics is a convenient technique to quantify spin-orbit torques, the present study shows that sufficient care should be taken concerning a possible phase variation along the stripe device, since otherwise a wrong estimation of spin-orbit torque efficiencies may result. Moreover, we have proposed a phase-independent and self-calibrated way to quantify the spin-orbit fields by using the shift of standing spin wave patterns excited by the combined action of current induced spin-orbit fields and Oersted field. This novel approach goes beyond the standard electrical measurements based on lineshape-analysis and solves a long standing problem in the determination of SOFs based on magnetization dynamics. Our method is not specific to Fe/GaAs, but can be also used for other systems, e.g., ferromagnetic metal/non-magnetic metal bi-layers.

**Methods:**

**Sample preparation.** To guarantee good interfacial quality, the Fe/GaAs (001) heterostructure is grown in a molecular-beam epitaxy (MBE) cluster. First, a 100-nm undoped GaAs buffer layer is deposited on GaAs (001) semi-insulating substrate at a substrate temperature of 600 ºC in a dedicated III-V MBE. After that, the wafer is transferred to another metal MBE without breaking the vacuum, and 0.8 nm or 3.5-nm-thick Fe is deposited epitaxially at room temperature. To avoid Fe oxidation, 2.5-nm thick Al capping layer is deposited in the same chamber, which is estimated to oxidize fully in air.

**Device.** For homogeneous excitation using a CPW, the Fe stripe with a dimension of 2.8 μm × 20 μm and with the long side along the [110]-orientation of GaAs (001) is integrated in the gap between the signal line and ground line. The CPW is designed to match the rf network which has an impedance of 50 Ω. The width of the signal line and the gap is 50 μm and 30 μm, respectively. The signal line and ground line are fabricated by 150 nm evaporated Au. For electric-current excitations, two kinds of devices with different widths are defined by electron-beam lithography and ion beam etching. For the phase variation experiment, devices with a size of 6.4 μm × 100 μm and with the long side along [110]-orientation of GaAs (001) have been used. For the determination of SOFs using SSW mode patterns, the device width is further narrowed down to 2.8 μm to enlarge the spacing of SSW modes. Devices along [110]-, [010]- and [$\bar{1}$10]-orientations have been fabricated to measure the symmetry of the SOFs.

**Measurements.** The measurements are carried out by time-resolved magneto-optical Kerr effect microscopy. To realize temporal resolution, the pulse train of a Ti:Sapphire laser (repetition rate of 80 MHz and pulse width of 150 fs) with wavelength of 800 nm is phase-locked to the microwave current $j_{rf}$, and a phase shifter is used to adjust the phase $\varphi_{l-j}$ between the laser pulse train and $j_{rf}$. The polar Kerr signal at a certain $\varphi_{l-j}$, $V_{Kerr}(\varphi_{l-j})$, is detected by a lock-in amplifier by phase modulating $j_{rf}$ at a frequency of 6.6 KHz. The microwave frequency is 12 GHz and the external magnetic field is applied along the long side of the stripes. The $V_{Kerr}(H, y)$ image is measured by scanning the device along $y$ direction through a piezo stage. All measurements are performed at room temperature.

**Acknowledgements**

This work is support by the German Science Foundation (DFG) via SFB 1277.

## Author contribution

L.C. and C.H.B. planned the study. R. I. and L.C. fabricated the devices, collected and analysed the data with the help from J.S and M.D.. M.K. D.S. and D.B. grew the samples. L.C., D.W. and C.H.B wrote the manuscript with input from all other co-authors. All authors discussed the results.

## Competing financial interests

The authors declare no competing financial interests.

## Data availability

The data that support the plots within this paper and other findings of this study are available from the corresponding author upon reasonable request.

## Figure captions

**Figure 1 | Schematic of device and driving fields. a**, Schematic of the device used for the detection of magnetization dynamics driven by electric-current. The out-of-plane component of the dynamic magnetization $m^z(t)$ is detected by time-resolved magneto-optical Kerr (TRMOKE) microscopy. A microwave current $j_{rf}^{FM}$ with a frequency $f$ of 12 GHz is fed into the Fe stripe deposited on a semi-insulating GaAs(001) substrate, and excites **m**(t) by the combination of spin-orbit field $h_{SOF}$ and Oersted field $h_{rf}^{FM,z}$. The external magnetic field $H$ is applied parallel to $j_{rf}^{FM}$. **b**, Phase relations in the TRMOKE set up. $\varphi_{l\text{-}j}$ is the phase difference between the laser pulse (green arrow) and $j_{rf}^{FM}$ (black

line), which is a controlled parameter. $\varphi_{j\text{-}m}$ is the assumed phase difference between $j_{\text{rf}}^{\text{FM}}$ and $m^z(t)$ (red dashed line). The phase difference between the laser pulse and $m^z(t)$, $\varphi_{l\text{-}m}$, is thus the sum of $\varphi_{l\text{-}j}$ and $\varphi_{j\text{-}m}$, i.e., $\varphi_{l\text{-}m} = \varphi_{l\text{-}j} + \varphi_{j\text{-}m}$. **c**, Current-orientation dependence of $h_{\text{SOF}}$ induced by Bychkov-Rashba-like (red) and Dresselhaus-like (green) spin-orbit interaction. Since $H$ is parallel to $j_{\text{rf}}^{\text{FM}}$, only the transverse components of $h_{\text{SOF}}$ excite magnetization dynamics. **d**, Lateral distributions of $h_{\text{SOF}}$ (red dashed line) and $h_{\text{rf}}^{\text{FM},z}$ (black line). $h_{\text{SOF}}$ is symmetrically distributed across $y$, while $h_{\text{rf}}^{\text{FM},z}$ is anti-symmetrically distributed. The different symmetries of the excitations lead to distinctive standing spin wave patterns, i.e., the symmetric $h_{\text{SOF}}$ excites even spin wave modes ($n = 1, 3…$), while the anti-symmetric $h_{\text{rf}}^{\text{FM},z}$ excites odd modes ($n = 2, 4 …$).

**Figure 2 | Determination of $\varphi_{j\text{-}m}$ for electric-current driven magnetization dynamics.** Position dependent Kerr voltage $V_{\text{Kerr}}$ measured at the center of the stripe ($y = 0$, where $h_{\text{rf}}^{\text{FM},z} = 0$) for **a** Fe thickness $t_{\text{Fe}} = 3.5$ nm and **b** $t_{\text{Fe}} = 0.8$ nm. For both devices, $\varphi_{l\text{-}j}$ is set to 50º and the device dimension is 6.4 μm × 100 μm. One can see that both the magnitude and the lineshape for $t_{\text{Fe}} = 3.5$ nm remain unchanged but change dramatically for $t_{\text{Fe}} = 0.8$ nm. The solid lines in **a** and **b** are fits to equation (3), which give the magnitude of $\varphi$. The bump at about 66 mT for $t_{\text{Fe}} = 3.5$ nm is due to the formation of a standing spin wave, see discussion in below. **c,** Position dependence of $\varphi_{j\text{-}m}$ obtained from equation (4), displaying a clear variation of $\varphi_{j\text{-}m}$ for $t_{\text{Fe}} = 0.8$ nm.

**Figure 3 | Eigenmodes and distribution of confined SSWs. a**, Calculated eigenmodes

for a laterally confined Fe/GaAs stripe with $w = 2.8$ μm. The external magnetic field $H$ is applied along the [110]-direction of GaAs, and the intersection defines the required $H_R$ for each standing spin wave (SSW) mode at $f = 12$ GHz. **b**, The lateral distribution of SSW modes for $n = 1$ - 5. The symmetric modes ($n = 1, 3$ and 5) can be excited by symmetric excitations; anti-symmetric modes ($n = 2$ and 4) can be excited by anti-symmetric excitations.

**Figure 4 | SSWs driven by a symmetric excitation. a** SSWs detected in a Fe/GaAs stripe by TR-MOKE microscopy, where the magnetization dynamics is excited by a homogeneous (symmetric) out-of-plane Oersted field through a coplanar waveguide (CPW). Only symmetric odd modes ($n = 1, 3$ and 5) can be observed. The inset shows the schematic of the CPW device, where the Fe stripe is integrated into the gap of the CPW (for details see Methods), and $H$ is applied along the long axis of the stripe, i.e. along the [110]-direction of GaAs. **b**, Micromagnetic simulation of the SSW modes using MuMax3, which reproduces the experimentally observed modes well. In the simulation, we use the same material parameters as for the calculation of the dispersion relation and we convolve the simulation with a Gaussian beam profile (see Supplementary Note 3). **c**, Horizontal line cut of the Kerr signal at the center of the stripe ($y = 0$). The three peaks can be fitted by symmetric Lorentzians, and the positions of the 1st, 3rd and 5th mode are indicated by red, green and blue circles, respectively. **d**, Vertical cut of modes for $n = 1, 3$ and 5. All the modes show symmetric profiles and can be well fitted by MuMax simulations.

**Figure 5 | SSWs driven by electric-current. a**, Image of the TR-MOKE signal measured at $\varphi_{l\text{-}j} = 90°$ and $\mathbf{j}_{rf}^{FM}$ // **H** // [110]. **b**, Profiles of the first three modes, i.e., vertical cuts along the dashed lines in **a**. The position of the maxima (minima) of $n = 1$ ($n = 3$) shifts away from the center of the stripe by $\Delta \sim 0.4$ µm due to the interference with the 2$^{nd}$ mode, as indicated by the dashed line. **c**, Corresponding image of $V_{\text{Kerr}}^{\varphi\text{-free}}$ for a [110]-device obtained by square and root operation of $V_{\text{Kerr}}(0°)$ and $V_{\text{Kerr}}(90°)$, i.e., $V_{\text{Kerr}}^{\varphi\text{-free}} = \sqrt{[V_{\text{Kerr}}(0°)]^2 + [V_{\text{Kerr}}(90°)]^2}$. **d**, Horizontal cut of $V_{\text{Kerr}}^{\varphi\text{-free}}$ at $y = 0$, which can be fitted by a symmetric Lorentzian.

**Figure 6 | Determination of SOF by the shape of SSW pattern.** Image of $V_{\text{Kerr}}^{\varphi\text{-free}}(H, y)$ signal for **a** $\mathbf{j}_{rf}^{FM}$ // **H** // [110], **b** $\mathbf{j}_{rf}^{FM}$ // **H** // [$\bar{1}$10] and **c** $\mathbf{j}_{rf}^{FM}$ // **H** // [010]. In the plots, $j_{rf}^{FM}$ has been normalized to $1 \times 10^{11}$ Am$^{-2}$. The configurations of the SOFs induced by Bychkov-Rashba-like $h_R$ and Dresselhaus-like $h_D$ are also presented in the insets. $h_R$ and $h_D$ are constructive for [110]-oriented devices, but destructive for [$\bar{1}$10]-oriented devices. For [010]-orientation, only $h_R$ is detected. Corresponding images obtained by micromagnetic simulations for devices oriented along **d** [110], **e** [$\bar{1}$10] and **f** [010].